\documentclass[RNAAS]{aastex62}
\usepackage{amsmath}
\usepackage{CJK}

\begin{document}
\begin{CJK*}{UTF8}{gbsn}

\title{Kepler-730b is Probably a Hot Jupiter with a Small Companion}

\correspondingauthor{Wei Zhu}
\email{weizhu@cita.utoronto.ca}

\author{Wei~Zhu (祝伟)}
\affiliation{Canadian Institute for Theoretical Astrophysics, University of Toronto, 60 St. George Street, Toronto, ON M5S 3H8, Canada}

\author{Fei Dai (戴飞)}
\affiliation{Department of Physics and Kavli Institute for Astrophysics and Space Research, Massachusetts Institute of Technology, Cambridge, MA, 02139, USA}
\affiliation{Department of Astrophysical Sciences, Princeton University, 4 Ivy Lane, Princeton, NJ, 08544, USA}

\author{Kento Masuda}
\affiliation{Department of Astrophysical Sciences, Princeton University, 4 Ivy Lane, Princeton, NJ, 08544, USA}
\affiliation{NASA Sagan Fellow}


\section{}

WASP-47 is so far the only confirmed system with a hot Jupiter and small, nearby companions \citep{Becker:2015}. This system challenged the high-eccentricity migration for hot Jupiter formation \citep[e.g.,][]{Dawson:2018}. To enable a systematic understanding of the prevalence of such systems, more detections are needed.

\begin{figure*}
\epsscale{1.}
\plottwo{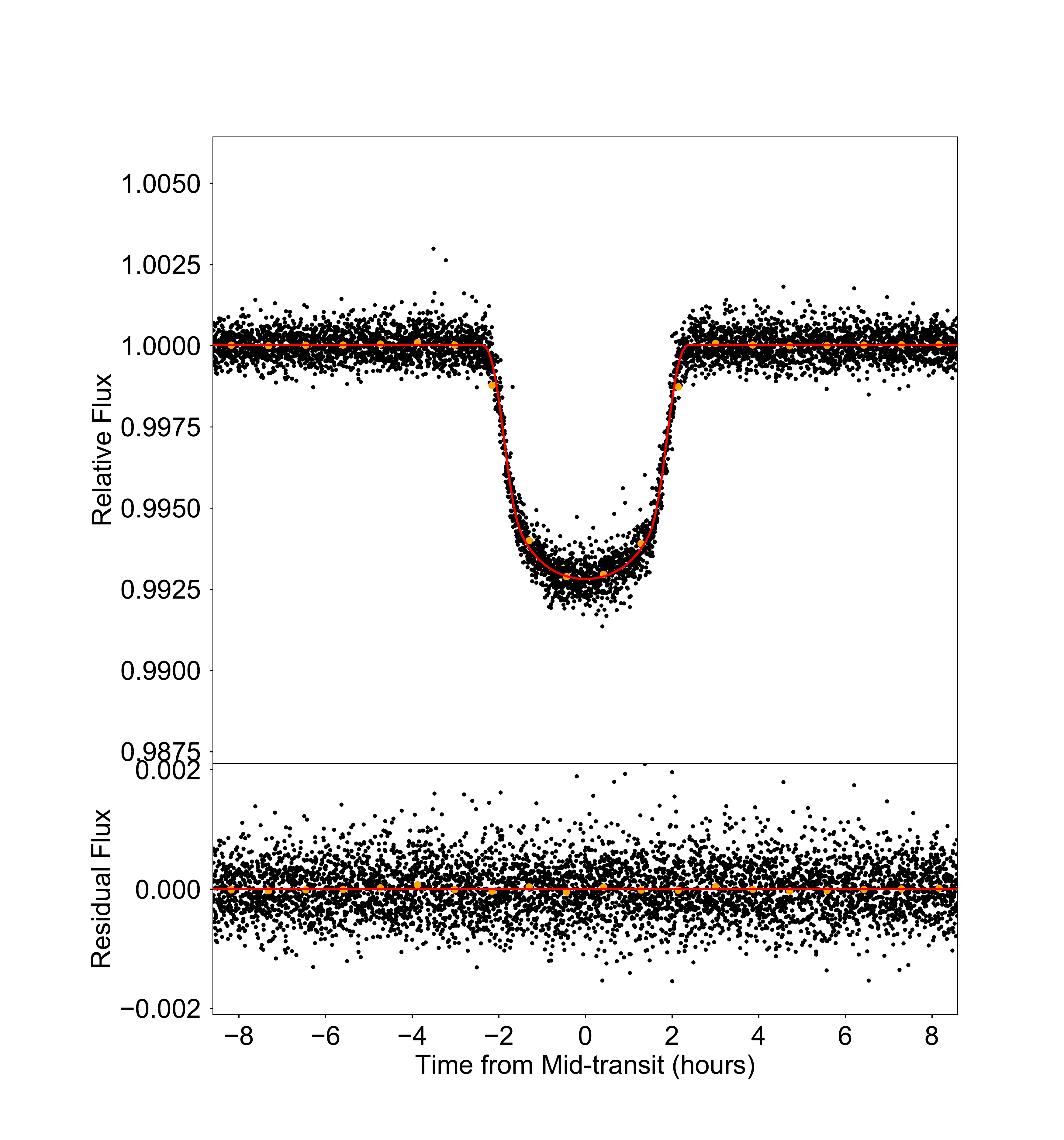}{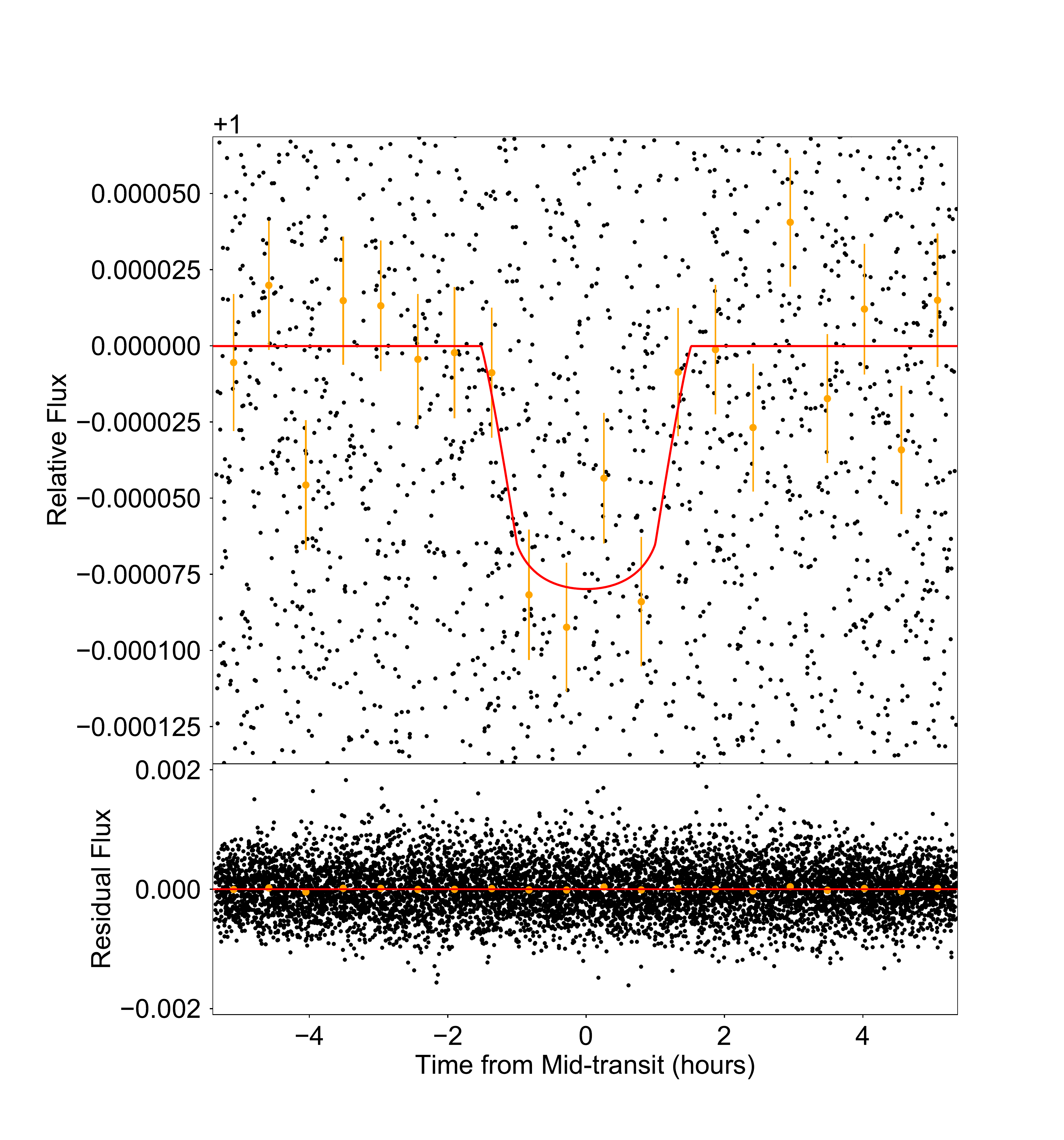}
\caption{Phase-folded transit light curve, best-fit model, and residuals of the hot Jupiter Kepler-730b (left panel) and the small companion KOI-929.02 (right panel). The black dots are the individual data points, and the orange bars are the binned data points.} \label{fig:lc}
\end{figure*}

Here we draw attention to a candidate system with one hot Jupiter and one small, nearby companion. The recent \emph{Kepler} data release \citep[DR25;][]{Thompson:2018} revealed a small inner companion to the already confirmed hot Jupiter Kepler-730b. With the updated Gaia stellar parameters from \citet{Berger:2018}, the host star (KOI-929, KIC 9141746) is an F-type main-sequence star with effective temperature $T_{\rm eff}=6126\pm214$ K and radius $1.30^{+0.13}_{-0.11}~R_\odot$, and has an apparent \emph{Kepler} magnitude $K_p=15.6$. Kepler-730b has updated radius $R_{\rm p}=11.36^{+1.14}_{-0.98}~R_\oplus$ and orbital period $P=6.492$ d, and the newly discovered companion, KOI-929.02, has $R_{\rm p}=1.45^{+0.15}_{-0.20}~R_\oplus$ and $P=2.852$ d. The phase-folded light curves of both transits are shown in Figure~\ref{fig:lc}. The small companion was detected at S/N$=10.1$, thus above the standard threshold (7.1), and it also passed all the validation test performed by the automated search pipeline. We refer to the \emph{Kepler} Candidate Overview Page
\footnote{\url{https://exoplanetarchive.ipac.caltech.edu/cgi-bin/DisplayOverview/nph-DisplayOverview?objname=KOI-929.02&type=KEPLER_CANDIDATE}}
under the NASA Exoplanet Archive \citep{Akeson:2013} for details of the detection, validation, and planetary parameters.

Although the small companion passed all the validation tests, there remains a possibility that this small companion transits a star that is different from the hot Jupiter host. In principle, such a scenario can be excluded if one can measure the centroid shift between in and out of transit in \emph{Kepler} images. However, such an effect is not measurable in the current case. The closest star (KIC 9141752, with $K_p=19.1$) is $5\farcs8$ away from the target position, and if this is the real host of the small companion, the transit depth should be $\sim0.0025$, and the expected centroid shift is small enough to remain undetected. Nevertheless, follow-up observations with 10-m telescopes can potentially resolve this issue by detecting (or not detecting) such a transit on the 19.1 magnitude star at the given ephemeris.

The transit timing variation (TTV) effect would be the ideal way to confirm the nature of such systems \citep[e.g.,][]{Becker:2015}. However, given the marginal detection of the small companion transit, TTV was not detected. Nevertheless, we find consistent mean stellar densities $\rho_\star=0.57\pm0.09$ g cm$^{-3}$ and $0.83\pm0.62$ g cm$^{-3}$ from modeling the light curves of the hot Jupiter and the small companion, respectively, under the reasonable assumption that both objects have (nearly) circular orbits. The two consistent densities agree with that the two companions indeed orbit around the same host. Furthermore, they are also more consistent with the density ($\rho_\star=0.65^{+0.17}_{-0.15}$ g cm$^{-3}$) of the assumed host (KIC 9141746) than the density ($\rho_\star=1.85^{+0.42}_{-0.55}$ g cm$^{-3}$) of the background star (KIC 9141752). These values are derived by the \texttt{isochrones} code \citep{Morton:2015} from fitting the Dartmouth stellar evolutionary models \citep{Dotter:2008} to observed colors from the \emph{Kepler} Input Catalog (\textit{grizJHK} for KIC 9141746 and \textit{gri} for KIC 9141752, corrected following \citealt{Pinsonneault:2012}) and parallax information from the Gaia DR2 \citep{Gaia:2018}. For the parallax values we adopted the distance estimates provided by \citet{BJ:2018}. We note that the parallax of KIC 9141752 is not well constrained and thus the result depends on the adopted distance prior.

To summarize, the KOI-929 system is probably another planetary system with hot Jupiter and small, nearby companion, after the outstanding WASP-47 system \citep{Becker:2015}, and so far the only such system (out of 46 reliable hot Jupiter systems in \citealt{Berger:2018}) in the prime \emph{Kepler} mission. The nature of this system, if confirmed, would suggest that hot Jupiters with small, nearby companions are probably more common than we used to believe. Detailed studies of such systems will help in resolving the puzzle of hot Jupiter origin. The ongoing Transiting Exoplanet Survey Satellite \citep[TESS;][]{Ricker:2014} mission can potentially discover more systems similar to KOI-929 and WASP-47.

\acknowledgements
We would like to thank Songhu Wang, Andy Gould and Subo Dong for useful discussions.


\end{CJK*}
\end{document}